# Transmission Performance Analysis of Digital Wire and Wireless Optical Links in Local and Wide Areas Optical Networks

**Abd El–Naser A. Mohamed[1], Mohamed M. E. El-Halawany[2]**
**Ahmed Nabih Zaki Rashed[3*], and Amina E. M. El-Nabawy[4]**
**[1,2,3,4]Electronics and Electrical Communication Engineering Department**
**Faculty of Electronic Engineering, Menouf 32951, Menoufia University, EGYPT**
**[1]E-mail: Abd_elnaser6@yahoo.com, [3*]E-mail: ahmed_733@yahoo.com**
**Tel.: +2 048-3660-617, Fax: +2 048-3660-617**

*Abstract*—**In the present paper, the transmission performance analysis of digital wire and wireless optical links in local and wide areas optical networks have been modeled and parametrically investigated over wide range of the affecting parameters. Moreover, we have analyzed the basic equations of the comparative study of the performance of digital fiber optic links with wire and wireless optical links. The development of optical wireless communication systems is accelerating as a high cost effective to wire fiber optic links. The optical wireless technology is used mostly in wide bandwidth data transmission applications. Finally, we have investigated the maximum transmission distance and data transmission bit rates that can be achieved within digital wire and wireless optical links for local and wide areas optical network applications.**

## I. INTRODUCTION AND BACKGROUND

Optical Wireless communication, also known as free-space optical (FSO), has emerged as a commercially viable alternative to RF and millimeter-wave wireless for reliable and rapid deployment of data and voice networks [1]. RF and millimeter-wave technologies allow rapid deployment of wireless networks with data rates from tens of Mb/s (point-to-multipoint) up to several hundred Mb/s (point-to-point). However, spectrum licensing issues and interference at unlicensed bands will limit their market penetration [2]. Though emerging license-free bands appear promising, they still have certain bandwidth and range limitations. Optical wireless can augment RF and millimeter-wave links with very high (>1 Gb/s) bandwidth. In fact, it is widely believed that optical wireless is best suited for multi-Gb/s communication. The biggest advantage of optical wireless communication is that an extremely narrow beam can be used. As a result, space loss could be virtually eliminated (<10 dB). But few vendors take advantage of this and use a wide beam to ensure enough signal is received on the detector even as the transceivers' pointing drift. This scheme is acceptable for low data rates, but becomes increasingly challenging at multi-Gb/s rates. Our approach has been to shift the burden from the communication system to a tracking system that keeps the pointing jitter/drift to less than 100 μrad. With such small residual jitter, sub-milliradian transmitted beam widths can be used. In so doing, the data communication part of the system is relatively simple and allows us to scale up to, and even beyond, 10 Gb/s. The main challenge for optical wireless is atmospheric attenuation. Attenuation as high as 300 dB/km in very heavy fog is occasionally observed in some locations around the world [3]. It is impossible to imagine a communication system that would tolerate hundreds of dB attenuation. Thus, either link distance and/or link availability has to be compromised. It is also obvious, that the more link margin could be allotted to the atmospheric attenuation, the better the compromise is. As a result, in the presence of severe atmospheric attenuation, an optical link with narrow beam and tracking has an advantage over a link without tracking [4]. Recent years have seen a wide spread adoption of optical technologies [5] in the core and metropolitan area networks. Wavelength Division Multiplexing (WDM) transmission systems can currently support Tb/s capacities. Next generation Fiber-to-the-Home (FTTH) access networks are expected to rely on Passive Optical Networks (PONs) in order to deliver reliable, multi-megabit rates to the buildings serviced by the network. Time Division Multiplexing PON (TDM/PON) and Wavelength Division Multiplexing PON (WDM/PON) may constitute a reliable alternative to the Active PON, where routing is done using a large Ethernet switch. However, as optical technologies are starting to migrate towards the access networks the cost factor is a vital issue [6] to the economic prospects of the investments. Unless significant progress is achieved in optical component integration in the near future, in terms of the scale of integration and functionality, the cost of the optoelectronic components is not expected to diminish in view of the light specifications placed by TDM/PON and WDM/PON. More importantly, if the existing duct availability is limited, one may expect large investment costs due to the enormous fiber roll out required. Free-Space Optics (FSO) is being considered as an attractive candidate in order to establish ultra high Gb/s wireless connections. FSO systems are classified as indoor (optical LANs) and outdoor systems.

FSO is sometimes referred to as optical wireless since it basically consists of transmitting the optical signal directly into the atmosphere without the use of an optical fiber. FSO systems have high bit rates (1Gb/s is already commercially available, while 10 Gb/s systems may soon appear). WDM technology may also provide a further increase in the aggregate transmission capacity exceeding 100 Gb/s. However, as light is no longer guided by the optical fiber, the performance of outdoor FSO systems is mainly limited by environmental factors. It is widely recognized that fog is the worst weather condition for FSO systems causing attenuation that might well exceed 100 dB/Km under heavy fog conditions. Atmospheric scintillation, i.e. the change of light intensity in time is also another limiting factor. The scintillations, caused by random, thermally induced fluctuations of the refractive index along the propagation path, result in bit error rate penalties and FSO systems are therefore designed to have a power margin. Another important issue is the misalignment caused by building sways [7] due to thermal expansion, wind sway and vibration. Using systems with larger beam divergence however may mitigate some of these effects. Automatic tracking techniques have been developed to deal with this problem.

In the present study, we have investigated the transmission performance analysis of digital wire and wireless optical links in local and wide areas optical network over wide range of the affecting parameters. Moreover, we have analyzed parametrically and numerically the maximum transmission distance and transmission bit rates that can be achieved within digital wire and wireless optical links for optical networks.

## II. SIMPLIFIED OPTICAL NETWORK ARCHITECTURE WITH WIRE AND WIRELESS OPTICAL LINKS

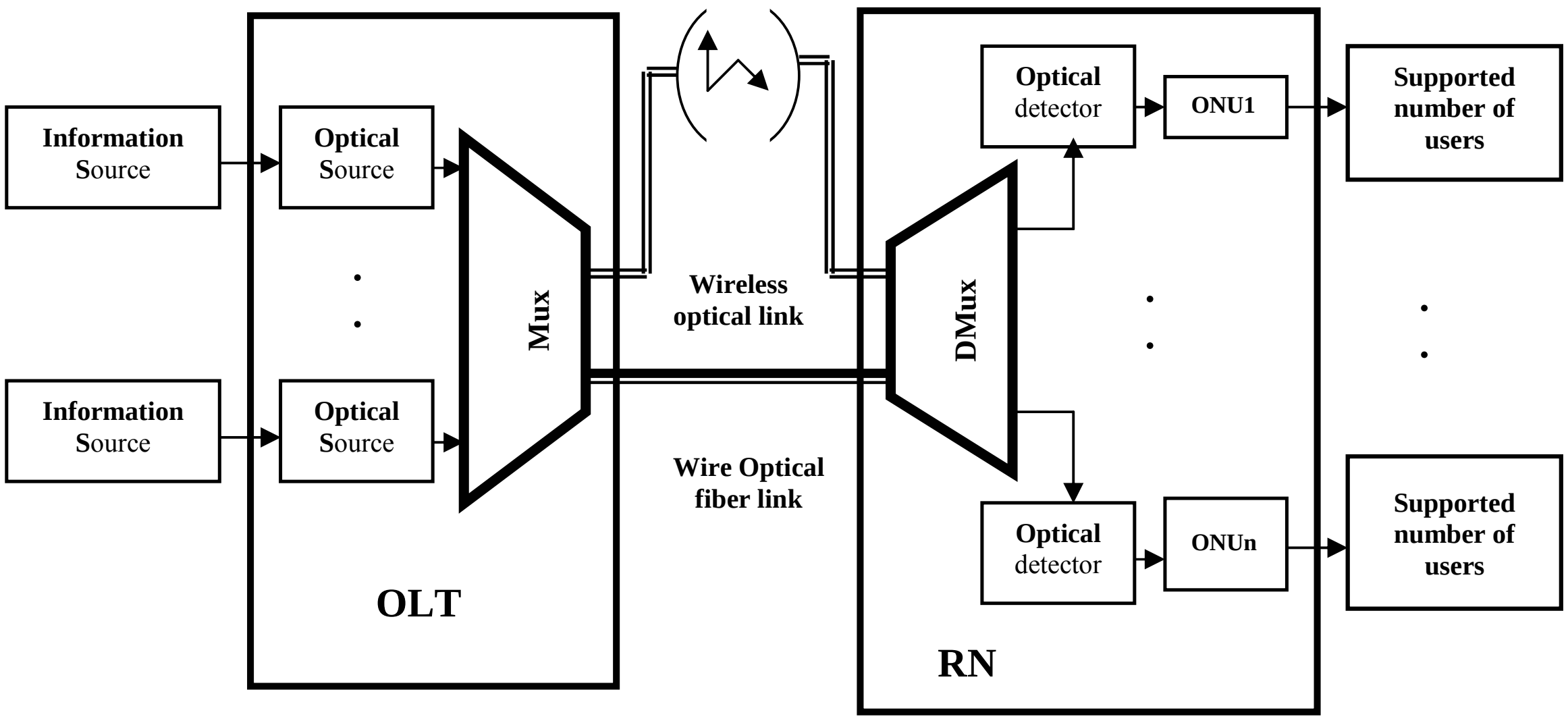

Figure 1. Simplified optical network Architecture Model with wire and wireless optical links.

The architecture model of passive optical network with different optical links is shown in Fig. 1. PON consists of many laser diodes as a source of optical signals which converts the electrical signal in the information source to optical signal, multiplexer (Mux) in the OLT, different optical fiber links, demultiplexer (Demux), optical network unit (ONU) in the remote node (RN), optical detector which converts the optical signal to electrical signal for processing to ONU and connects to the supported number of users. In the transmission direction, the information source (electrical signal) is transmitted from the backbone network to the OLT and according to different users and location, optical source [laser diode or light emitting diode] convert it in to optical signal and is transmitted into corresponding wavelength and multiplexed by Mux. When traffic arrives at RN, wavelengths are demultiplexed by Demux and sent to optical detector [Avalanch photodiode or PIN photodiode] convert the optical signal into electrical signal and then sent to ONUs which is distributed to different number of supported users. Wavelength division multiplexing passive optical networks has been regarded as a promising technology to meet the demands of various customers or most supported subscribers which include many kinds of broadband data such as a high speed internet, high data transmission data rate wireless transmission, and a real time video service. Recently, various techniques of access networks have been presented to increase transmission capacity, transmission distance and reduce the cost per supported user.

### *II. 1. Simplified basic configuration of digital wireless optical communication system*

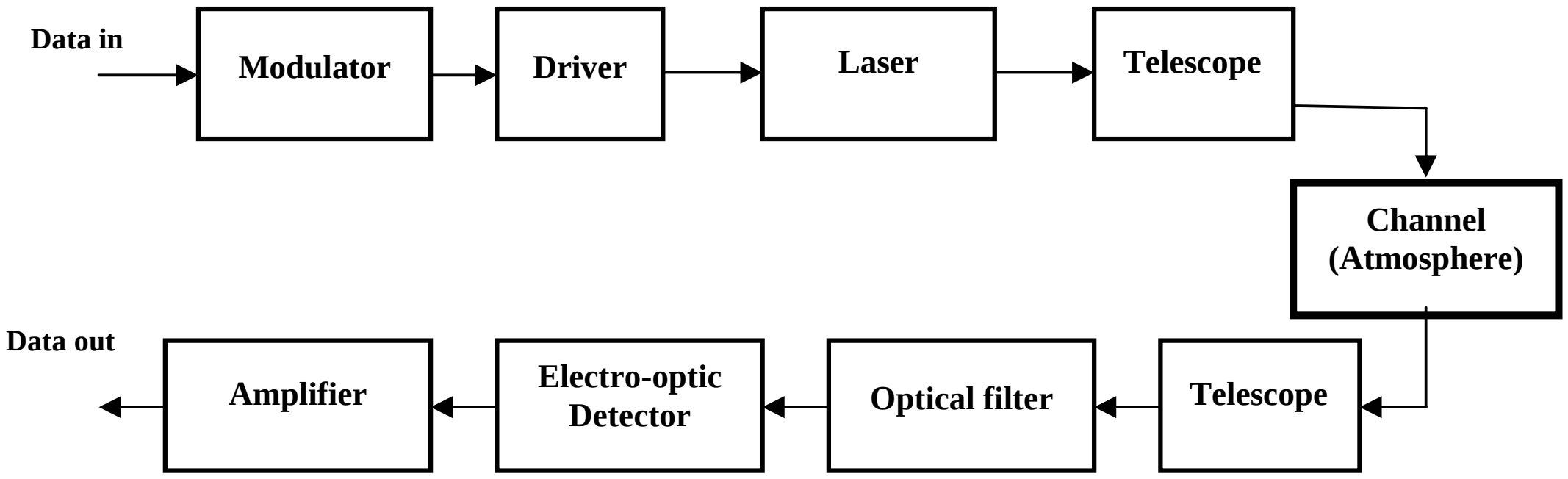

Figure 2. Basic configuration of wireless optical communication system.

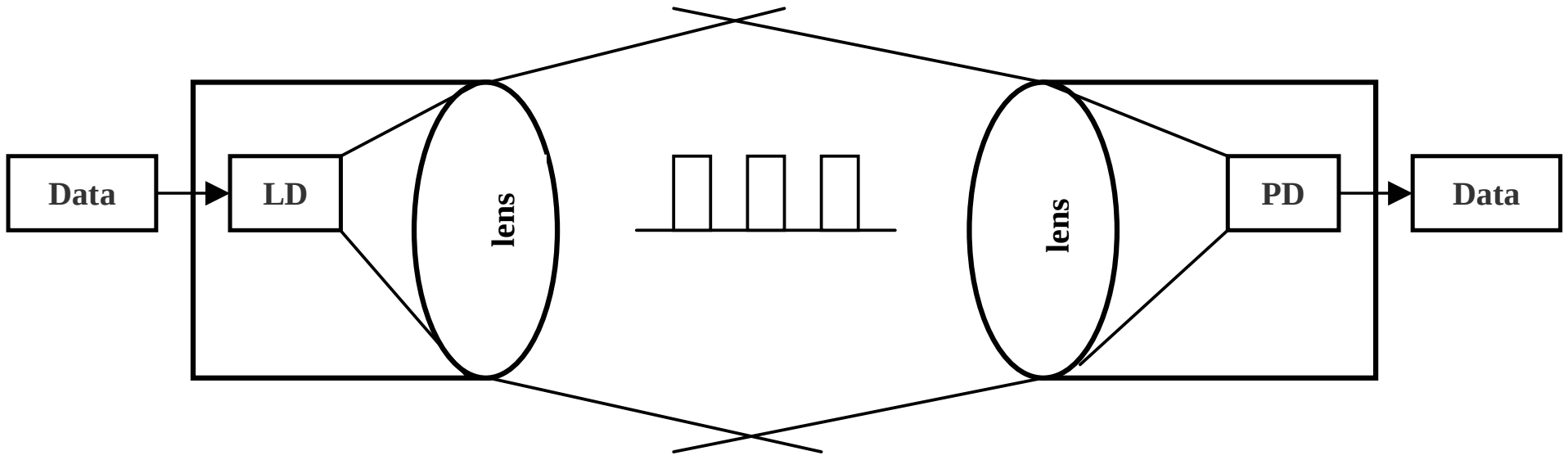

Figure 3. Schematic view of the configuration of wireless optical link.

Figures (2, 3) show the basic configuration of wireless optical communication system link. The link consists of an electrical/optical (E/O) conversion device and an optical/electrical conversion (O/E) device. The E/O conversion is accomplished by either the laser diode or the external modulator, while the O/E conversion by the photodiode such as PIN diode and APD. The transmission data rate is dependent on the modulation speed of the E/O devices. The wavelength division multiplexed (WDM) technology can increase the transmission capacity using a number of laser diodes and photodiodes with multiplexers. The wireless optical link is used for point-to-point applications such as the access link between a hub station and a subscriber terminal. The important parameter for the wireless optical communication link is an optical wavelength. The short wavelength in the range of 0.78-0.8 μm was first introduced to transmit a lower data rate. The long wavelength which is used for the fiber optic systems is in the range of 1.3-1.5 μm [8]. The advantage of the long wavelength is that the optical amplifiers are now available, and because of the amplification of optical carrier the transmission distance between a hub station and a subscriber terminal can be increased. The output power of the link should be designed by taken into account of the eye safety. Optical fiber maintenance is a very important issue to be consider in developing a high quality and reliable passive optical network [9].

### *II. 2. Simplified basic configuration of digital wire optical cable link*

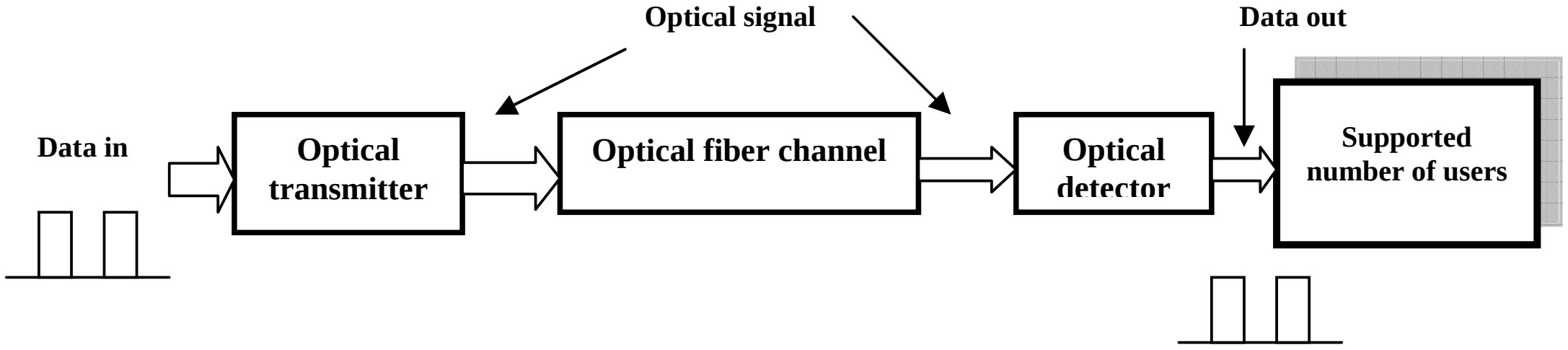

Figure 4. Basic configuration of digital optical link.

As shown in Fig. 4, the basic architecture view of the configuration of the digital optical link. Digital communications systems have many advantages over analogue systems brought about by the need to detect only the presence or absence of a pulse rather than measure the absolute pulse shape. Such a decision can be made with reasonable accuracy even if the pulses are distorted and noisy. For single wavelength systems, repeaters allow new clean pulses to be generated if required, preventing the accumulation of distortion and noise along the path. In optical communications systems, the pulse sequence is formed by turning on and off an optical source either directly or using an external modulator. The presence of a light pulse would correspond to a binary 1 and the absence to a binary 0. The two commonly used techniques for representing the digital pulse train are non return to zero (NRZ) and return to zero (RZ). In the case of NRZ, the duration of each pulse is equal to twice the duration of the equivalent RZ pulse. The choice of scheme depends on several factors such as synchronisation, drift etc. An ac coupled photoreceiver will generally not pass a signal with long sequences of '1's or '0's and so some form of RZ coding scheme would be required [10].

## III. BASIC SYSTEM MODEL AND EQUATIONS ANALYSIS

There are several important system issues that need to be considered in the theoretical model equations analysis of such an arrangement of digital wire or wireless optical cable links:

i) Optical signal wavelength: Most installed fibre is designed for use at 1.3 μm. If long haul links are necessary and multiple wavelength channels are needed then the wavelength must be in the 1.55 μm region. Single channel links can be implemented with a 1.3 μm system using optical repeaters to extend the reach.

ii) Digital wireless: wireless modulation avoids the requirement for samplers and digitisers at each telescope site allowing these to be situated at the correlator. Digital systems are much less prone to noise and non-linear effects and so can offer better quality signals over larger distances.

iii) Length of link: This will have an impact on the modulation technique used and the need for mid-span optical amplifiers (or electrical repeaters for single wavelength channels).

iv) Data rate: If a digital implementation is chosen, the data rate will define the maximum instantaneous bandwidth and hence the sensitivity of the radio astronomy measurement. Commercial equipment will require standard data rates to be used (2.5Gbps, 10Gbps) which may not be compatible with the radio astronomy front and back end.

### *III. 1. Wireless optical link design*

In the design of wireless optical link system, it is important to determine the link budget equation. The general link budget equation is given by [11]:

$$P_{received} = P_{transmit} \cdot \frac{57.295\, A_{receiver}}{(\theta L)^2} e^{-\alpha L} \quad (1)$$

where $p_{received}$ is the power at receiver (watt), $P_{transmit}$ is the transmission power (watt), $A_{receiver}$ is the receiver effective area ($m^2$), θ is the beam divergence (degrees), L is the length of the optical link (m), and α is the atmosphere absorption (dB/Km). The total loss coefficient is determined by:

$$\sigma L = \sigma_{rain} L + \sigma_{fog} L + \sigma_{snow} L + \sigma_{sc\,\mathrm{int}\,illation} \quad (2)$$

where $\sigma_{rain}$ is the absorption due to rain ($Km^{-1}$), $\sigma_{fog}$ is the absorption due to fog ($Km^{-1}$), $\sigma_{snow}$ is the absorption due to snow ($Km^{-1}$), and $\sigma_{scin}$ is the absorption due to scintillation ($Km^{-1}$). A variety of models exist for the calculation of these absorption coefficients. In the case of fog, the Kruse model according to:

$$\sigma_{fog}\,(Km^{-1}) = \frac{3.912}{V}\left(\frac{\lambda}{\lambda_0}\right)^{-q} \quad (3)$$

where V is the visibility at ($\lambda=\lambda_0$), Km, λ is the actual wavelength of the beam, μm, $\lambda_0$ is the reference wavelength in μm for the calculation of V, and the exponent q is the size distribution of the scattering particles and is equal to 1.3 if 6 Km < V < 50 Km, and equal to 0.585 $V^{1/3}$for low visibility V < 6 Km. Also to calculate the optical losses due to snow, the empiricial formula can be used:

$$\sigma_{snow}(dB/Km) = A S^b \quad (4)$$

where S is the snow fall rate (in mm/hour), A=5.42x$10^{-5}$ λ+ 5.9458, and b= 1.38. In the same way, to calculate the optical losses due to rain, the empiricial formula can be used:

$$\sigma_{rain}(dB/Km) = 1.076\, R^{2/3} \quad (5)$$

where R is the rain fall rate measure (in mm/hour). Finally the optical loss due to scitillation is calculated using the following expression [11]:

$$\sigma_{sc}^2 = 4.\left(23.17\left(\frac{2\pi}{\lambda}10^9\right)^{7/6}\right) C_n^2\, L^{11/6} \quad (6)$$

where $C_n^2$ is the scintillation strength (in $m^{-2/3}$). It should be noted that the case of wireless optical link system, fog induced absorption is the most impairment and can be significantly affect the performance of the system. A link budget for wireless optical link using one lens in the transmitter and one lens in the receiver is calculated. Different kind of losses are calculated that may cause power losses during transmission [11]. The factors that cause the majority of the losses for the system are the atmosphere attenuation and ray losses.

Equation (7) shows that the ray losses of the system depend on the radius of the receiver lens and the beam radius at the receiver unit. A Gussian beam intensity distribution is assumed [12]:

$$F_S = 10\log\frac{P_{receiver}}{P_{total}} = 10\log\left(1 - e^{-\frac{2R^2}{w\,(L)}}\right) \quad (7)$$

where L is the link distance, Km, $F_S$ is the ray losses, dB, $P_{total}$ is the total beam power at L, watt, R is the lens radius ,mm, w (L) is the beam radius, mm. Geometrical losses occur due to the diverence of the optical beam. These losses can be calculated using the following formula [12]:

$$\frac{A_R}{A_T} = \left( \frac{57.295\, D_R}{D_T + 100.\ d.\theta} \right)^2 , \qquad (8)$$

where $A_R$ is the effective area of the receiver lens, $A_T$ is the effective area of the transmitter lens, $D_R$ is the diameter of the transmitting lens, $D_T$ is the diameter of the receiving lens, d is the distance between the wireless optical transmitter and receiver, θ is the divergence of the transmitted laser beam in degrees. Based on curve fitting Matlab Program, the fitting equations between optical signal to noise ratio (OSNR), the operting signal wavelength for transmitter and receiver, and the wireless optical link length are [13]:

$$OSNR = 17.35 - 12.27\, L + 7.05\, L^2 - 5.87\, L^3 \quad , \qquad (9)$$

$$OSNR = 3.85 - 10.73\, \lambda + 2.13\, \lambda^2 + 9.75\, \lambda^3 \quad , \qquad (10)$$

The radio frequency transmission response provide the relative loss or gain in a wireless communication system links with respect to the signal frequency. Any signal attenuation due to the wireless communication links can be expressed as follows [12]:

$$Transmission(dB) = 10 \log \left( \frac{P_{transmitter}}{P_{incident}} \right) , \qquad (11)$$

where $P_{transmitter}$ is the radio frequency power calculated at the output of the receiver, and $P_{incident}$ is the radio frequency power calculated at the input to the laser transmitter. Based on curve fitting Matlab Program, the fitting equations between transmission response, operating radio frequency, and amplification range are [12]:

$$Transmission(dB) = 10.82 - 2.05\, f + 7.42\, f^2 - 4.23\, f^3$$

(without amplification), (12)

$$Transmission(dB) = 3.09 + 13.65\, f - 2.56\, f^2 + 1.85\, f^3$$

(with amplification) (13)

The Shannon capacity theorem to calculate the maximum data transmission bit rate or the maximum channel capacity for the wireless optical links is as follows:

$$C = B.W \log_2 (1 + OSNR), \quad bits/\sec \qquad (14)$$

### *III. 2. Digital wire optical cable link design*

Digital communications systems have many advantages over analogue systems brought about by the need to detect only the presence or absence of a pulse rather than measure the absolute pulse shape. Such a decision can be made with reasonable accuracy even if the pulses are distorted and noisy. For single wavelength systems, repeaters allow new clean pulses to be generated if required, preventing the accumulation of distortion and noise along the path. Chromatic dispersion is caused by a variation in group velocity in a fibre with changes in optical frequency. The set of pulses generated by a laser which by virtue of the laser linewidth and signal modulation contains a spectrum of wavelengths. As it traverses the fibre, the shorter wavelength components travel faster than the longer wavelength components and as a result, each pulse experiences broadening. By the time the pulses reach the receiver, they may have broadened over several bit periods and be a source of errors (inter symbol interference). The measure of chromatic dispersion is D, in units of psec/nm.km, which is the amount of broadening in picoseconds that would occur in a pulse with a bandwidth of 1nm while propagating through 1km of fibre. The chromatic dispersion factor is given by [13]:

$$\gamma = \frac{\lambda}{\pi\, c} B^2\, L\, D \ , \qquad (15)$$

where B is the data rate, L is the fiber path length, and c is the speed of light in a vacuum. As the optical components propagate through the fibre, the inherent birefringence causes one of the components to be delayed with respect to the other. In high bit rate systems, this differential group delay can lead to signal distortions and hence a degradation in the BER of the received signal. The group delay between two polarisation components is called the differential group delay, Δτ. Its average is the Polarization mode dispersion (PMD) delay in psec and is expressed by the PMD coefficient in $ps/km^{1/2}$. The PMD does not increase linearly, but with the square root of transmission distance.

$$\Delta\tau = \sqrt{L}\, .\Delta\tau_{coeff} \ , \qquad (16)$$

where L is the transmission distance, and $\Delta\tau_{coeff}$ is the PMD coefficient. Taking into account the statistical character of PMD variations, if a 1 dB power pentaly due to PMD can be accepted then:

$$\Delta\tau_{\max} \le \frac{T}{10} \quad , \qquad (17)$$

where T is the bit period. Setting T as $1/B_0$ we obtain:

$$L \le \frac{1}{100\, .\, B_0^2\, .\, \Delta\tau_{coeff}^2} \quad , \qquad (18)$$

where $B_0$ is the bit rate. The receiver sensitivity is defined as the minimum number of photons per bit necessary to guarantee that the bit error rate (BER) is smaller than $10^{-9}$ . This sensitivity corresponds to an optical energy $h\nu\, n_0$ and an optical received power as follows:

$$P_{receiver} = h\upsilon\, n_0\, B_0 \ , \qquad (19)$$

This power is proportional to the total bit rate $B_0$. In a saturated attenuation-limited link, the link budget in dBm units as follows [14, 15]:

$$P_{receiver} = P_s - P_c - P_m - \alpha\, L, \ dB/km \qquad (20)$$

where $P_s$ is the source power, $P_m$ is the modulator power, α is the fiber loss in dB/Km, $P_c$ is the coupling loss, and L is the fiber length. When $P_{receiver}$ is converted to dB, it is evident that $P_{receiver}$ increases logarithmically with the data rate $B_0$. Therefore as the bit rate increases the power required to maintain the desired BER also increases. With this in mind, the derived maximum length of the digital optical link [15]:

$$L = \frac{1}{\alpha} \left( P_s - P_c - P_m - 10 \log \frac{n_0\, h\, \nu\, B_0}{10^{-3}} \right) \ , Km \qquad (21)$$

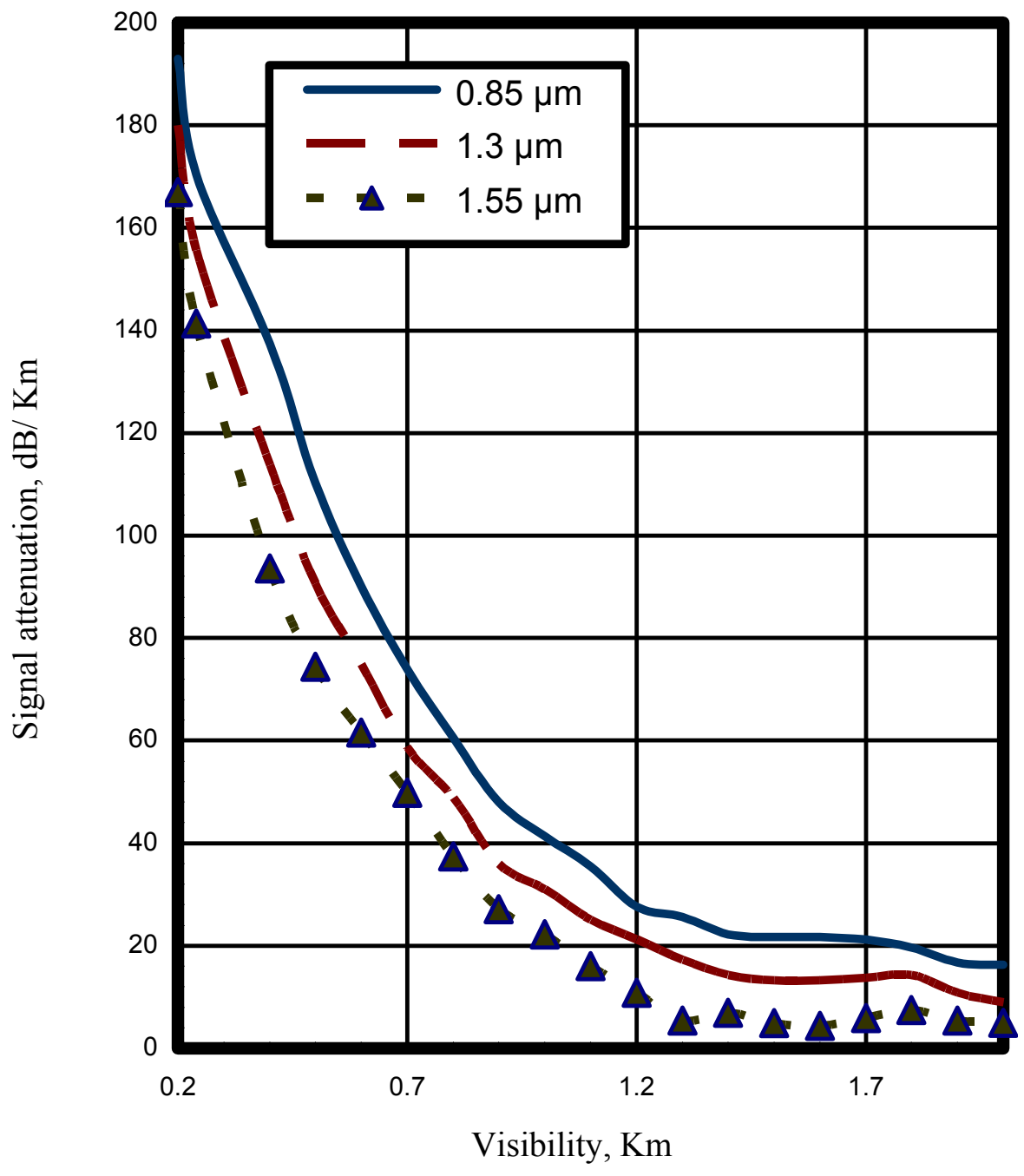

Figure 5. Variations of the signal attenuation with visibility for different laser diode wavelengths at the assumed set of parameters.

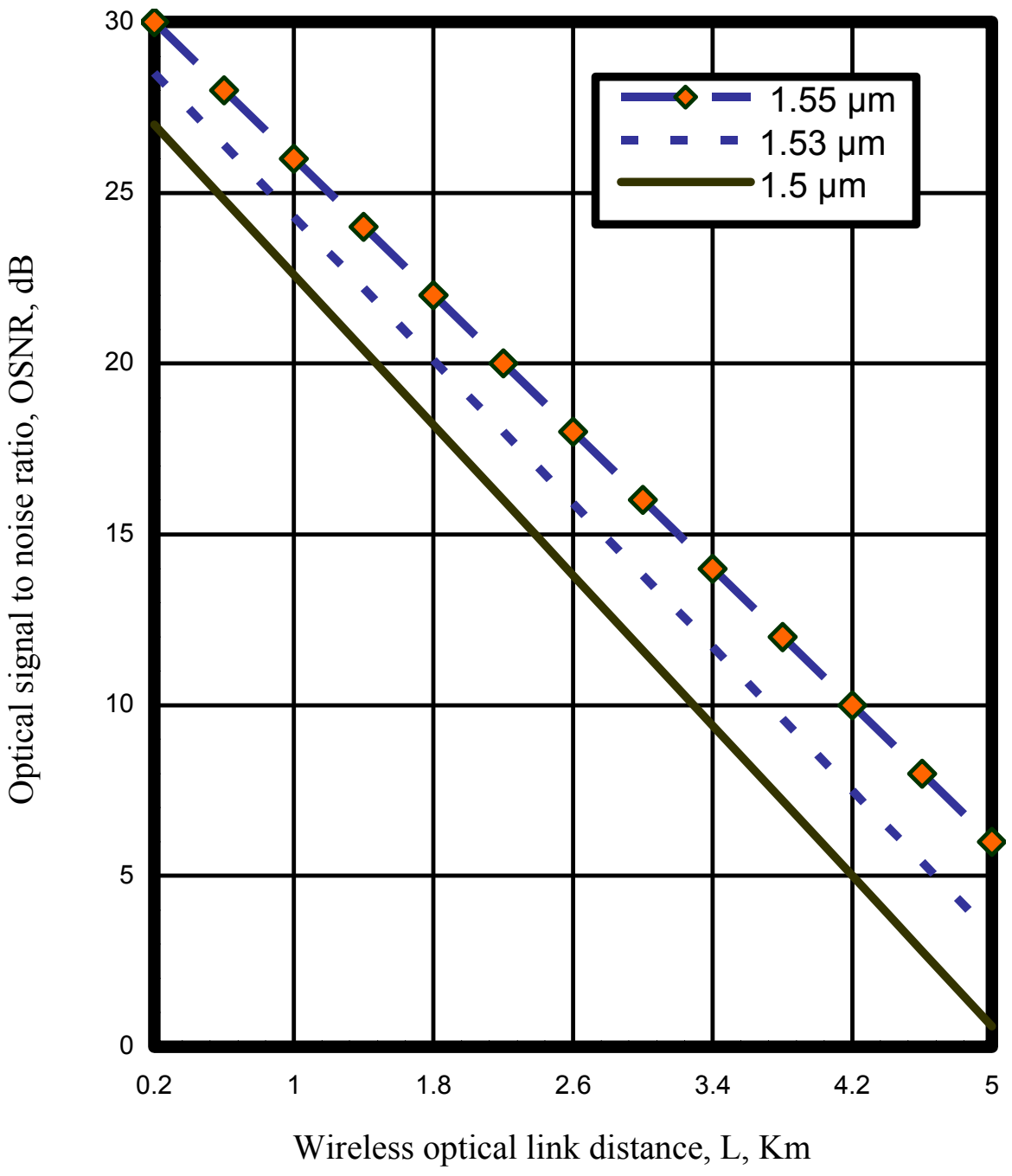

Figure 7. Variations of optical signal to noise ratio with wireless optical link distance at the assumed set of parameters.

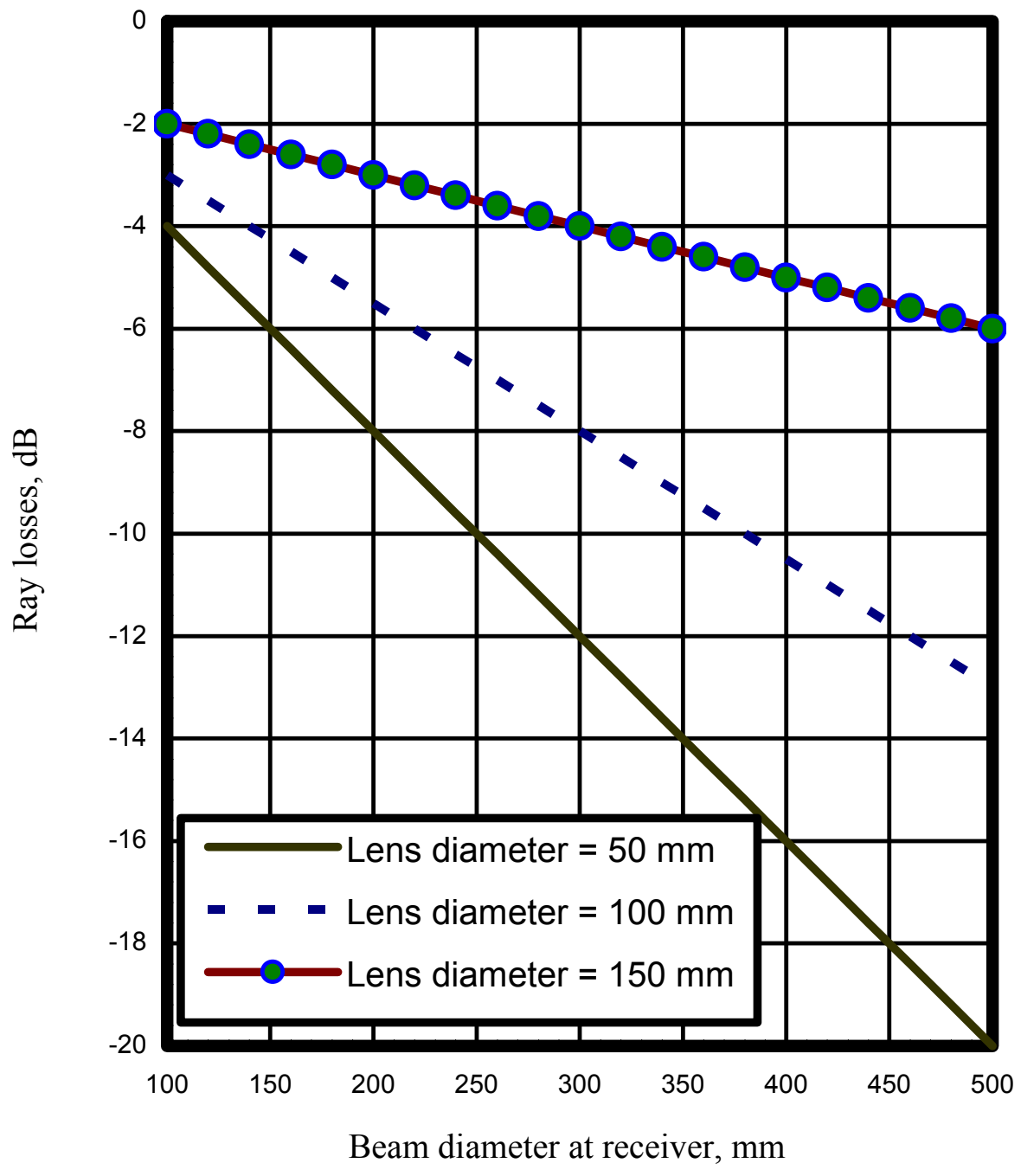

Figure 6. Variations of the ray losses with beam diameter at receiver for different lens diameter at the assumed set of parameters.

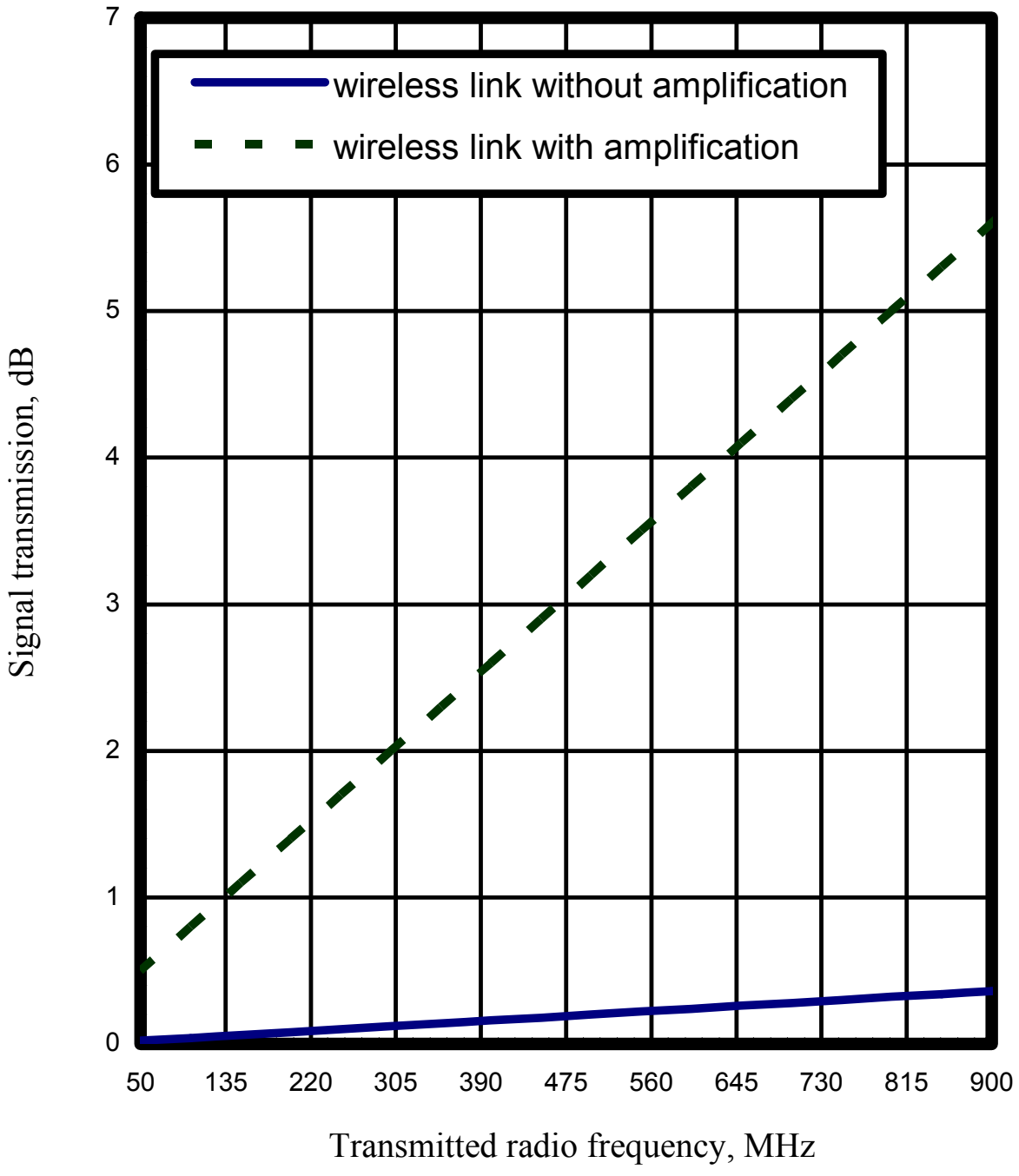

Figure 8. Variations of wireless transmission with transmitted radio frequency at the assumed set of parameters.

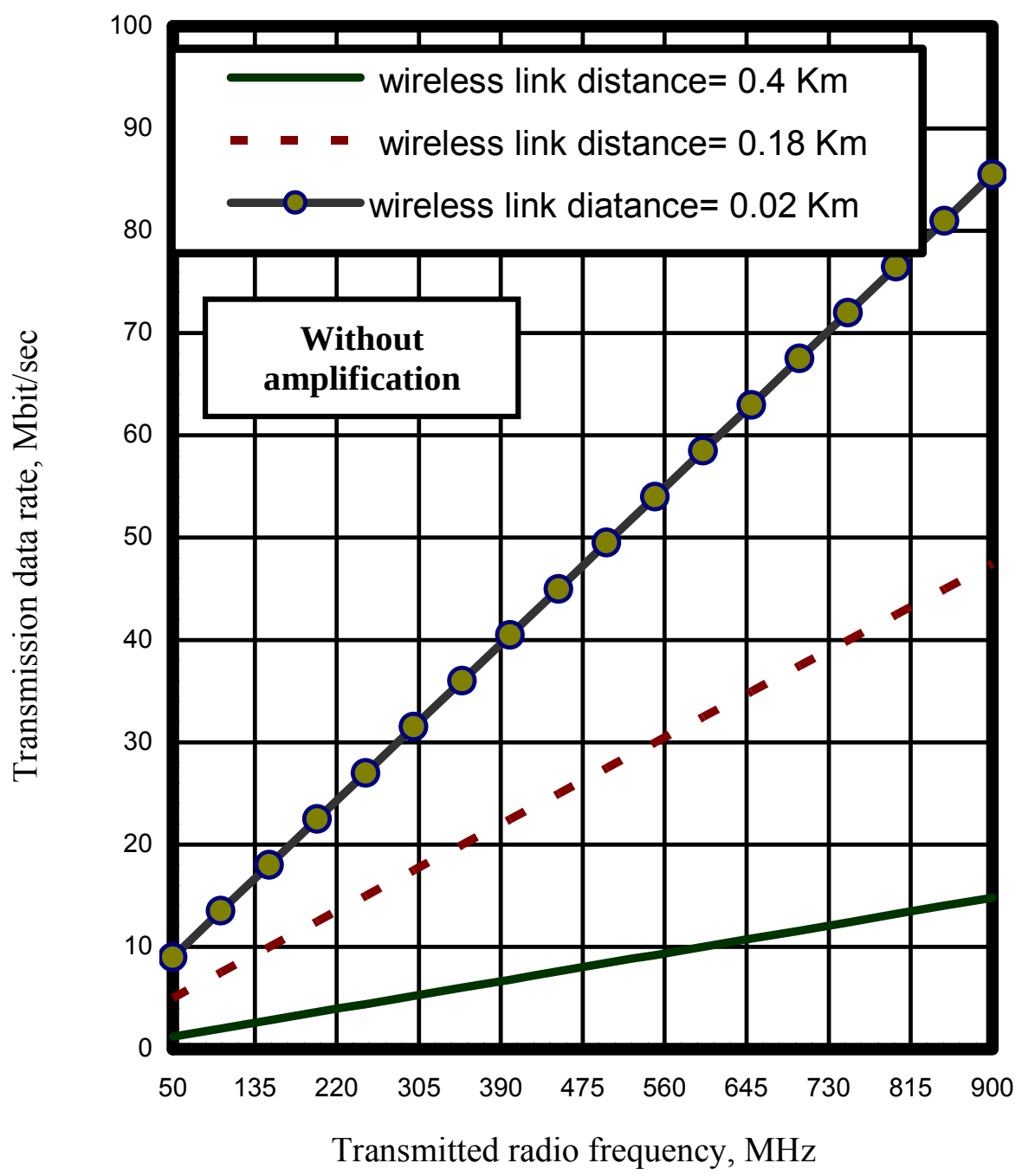

Figure 9. Variations of transmission data rate with transmitted radio frequency at the assumed set of parameters.

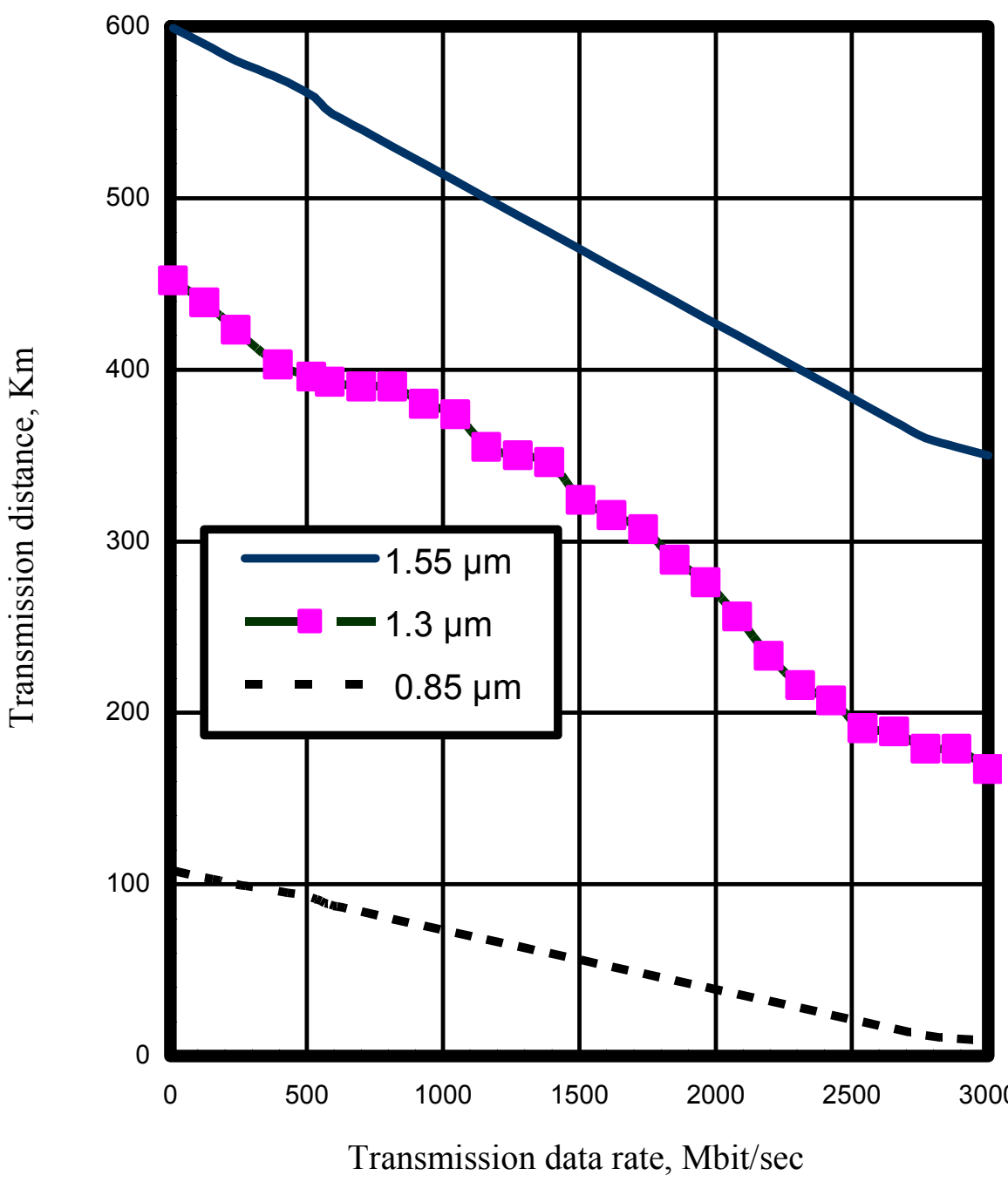

Figure 11. Variations of the transmission distance for digital optical link with transmission data rate at the assumed set of parameters.

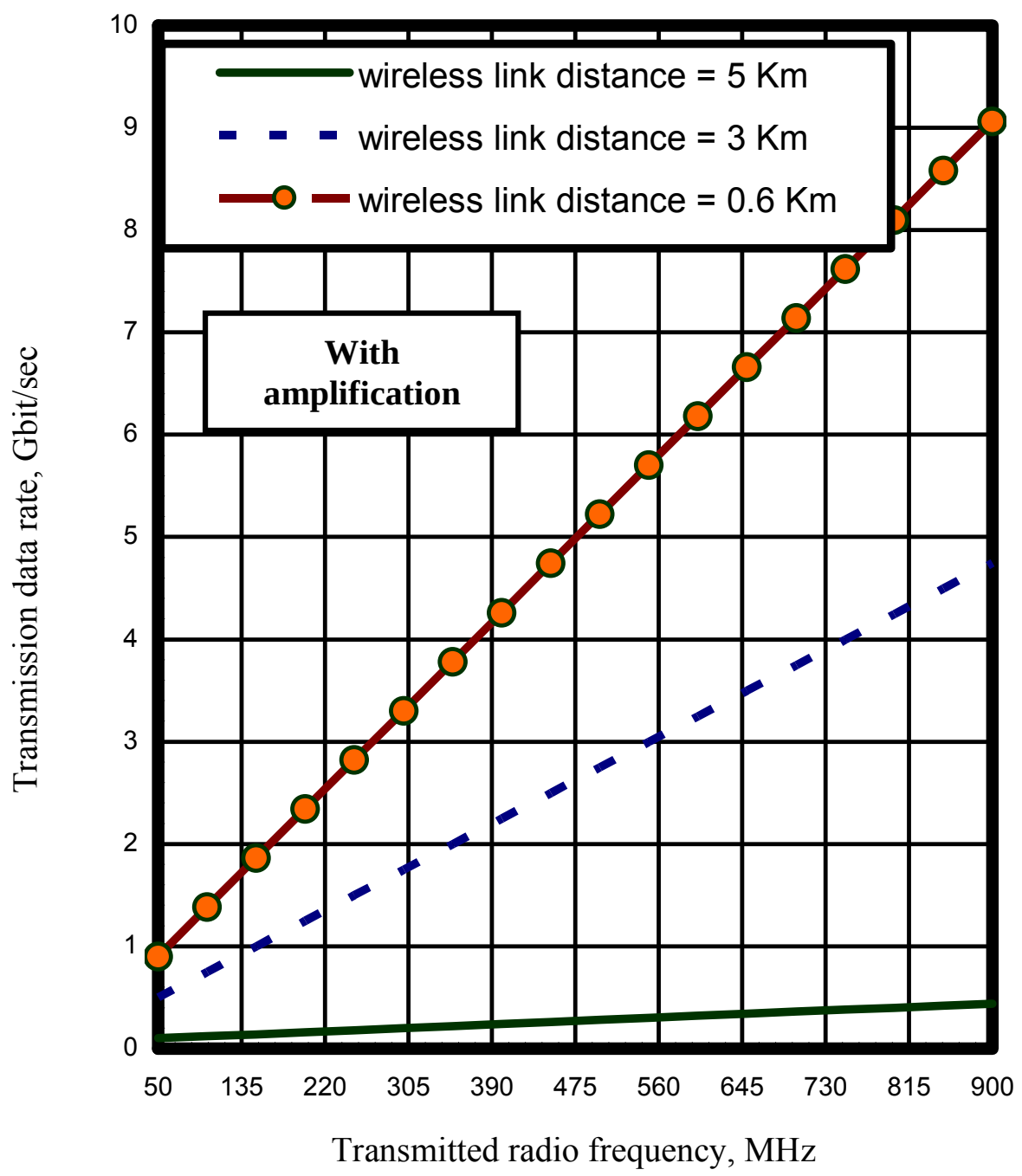

Figure 10. Variations of transmission data rate with transmitted radio frequency at the assumed set of parameters.

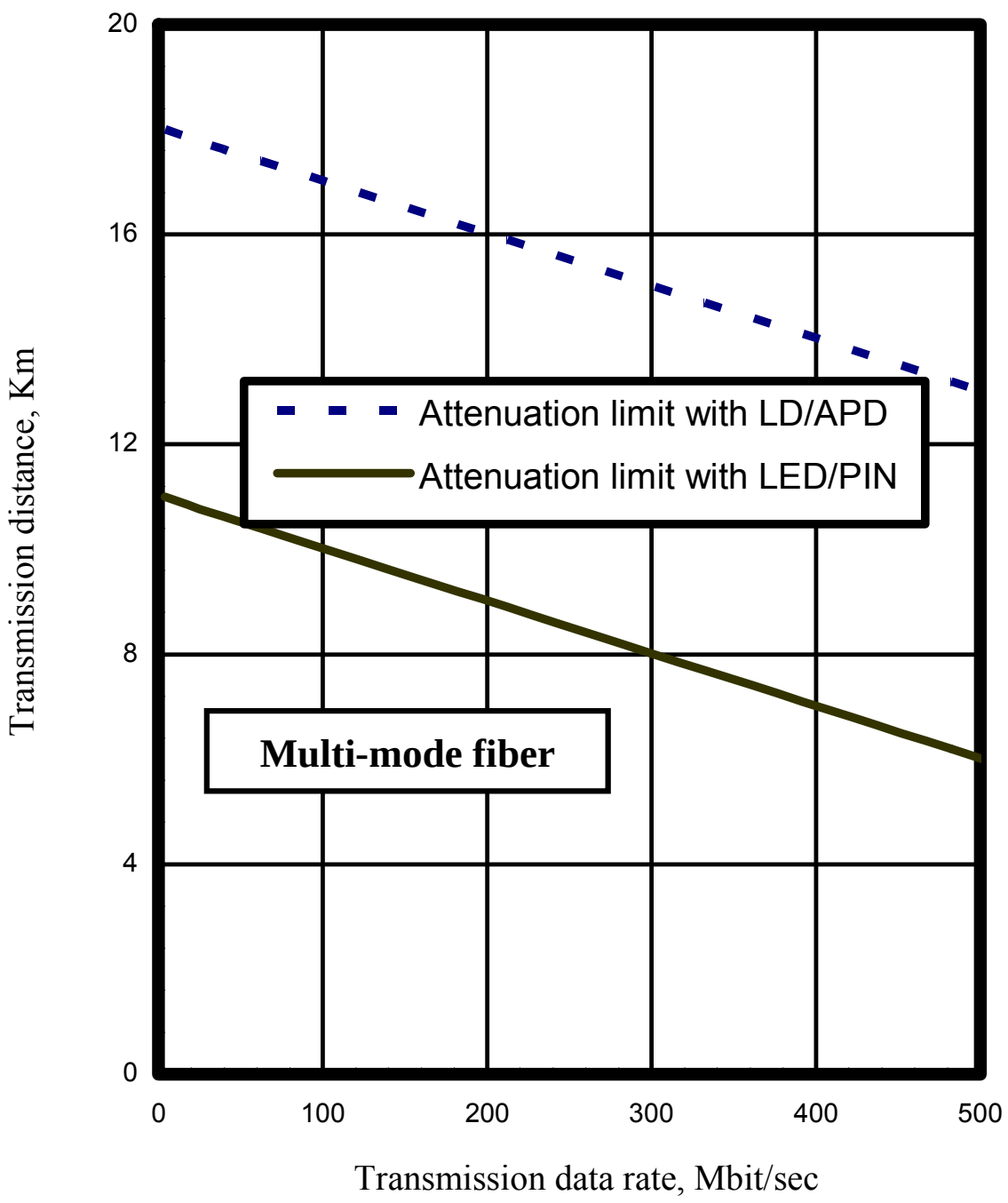

Figure 12. Variations of the transmission distance for digital optical link with transmission data rate at the assumed set of parameters.

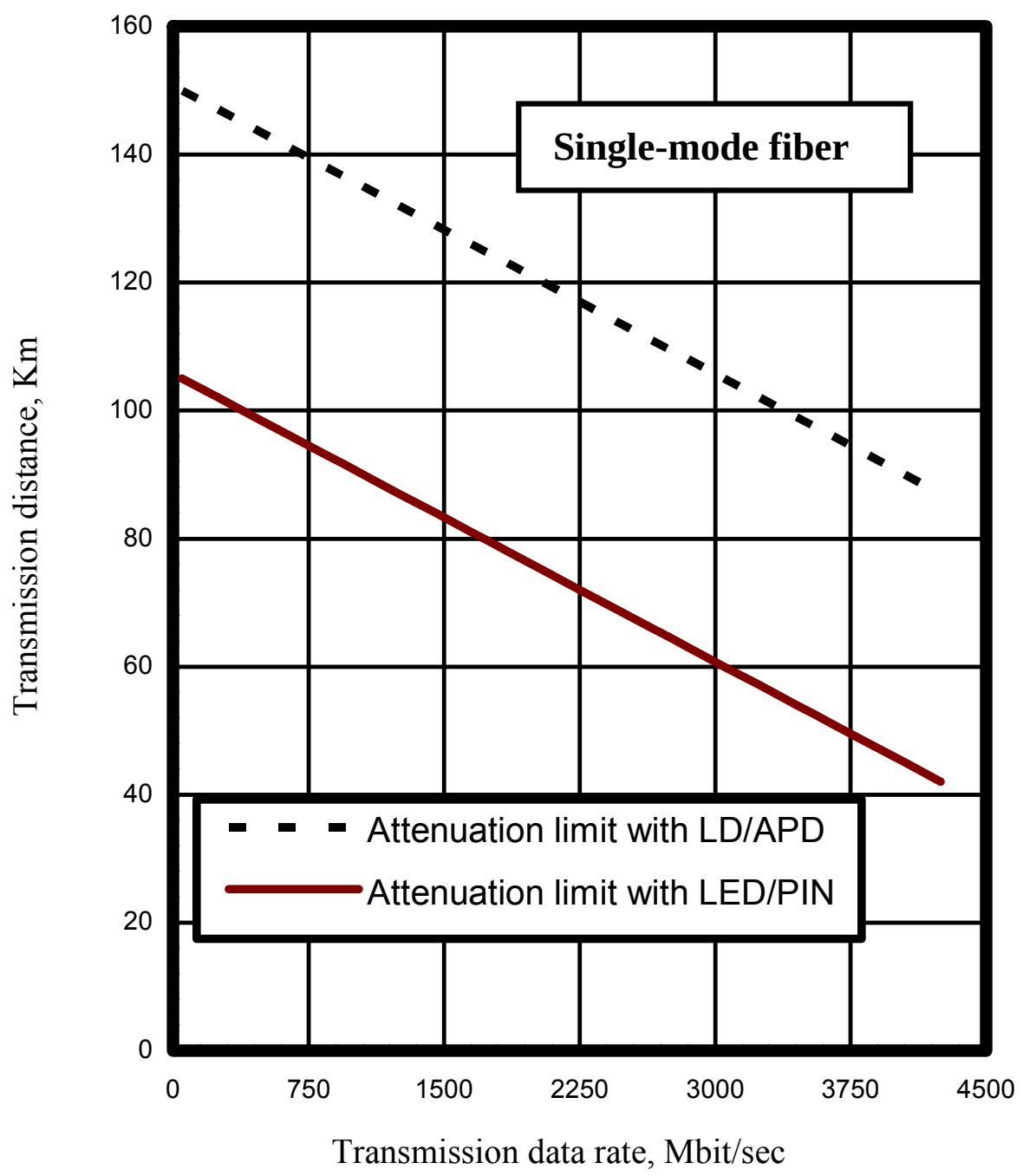

Figure 13. Variations of the transmission distance for digital optical link with transmission data rate at the assumed set of parameters.

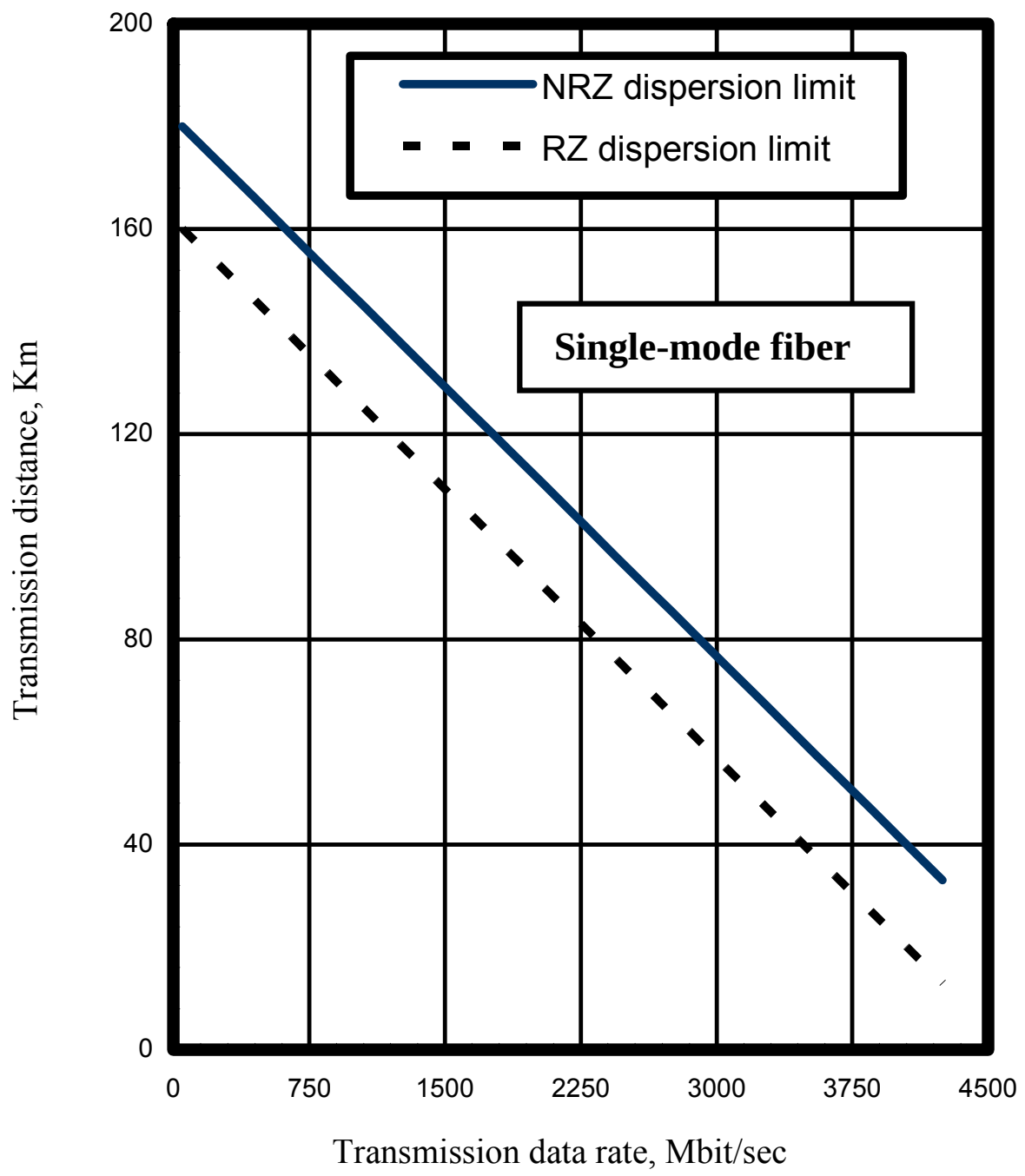

Figure 14. Variations of the transmission distance for digital optical link with transmission data rate at the assumed set of parameters.

The total rise time depends on: transmitter rise time ($t_{tx}$), group velocity dispersion ($t_{GVD}$), modal dispersion rise time ($t_{mod}$), and receiver rise time ($t_{rx}$), therefore the total rise time, $t_{sys}$, for the system is [15]:

$$t_{sys} = \left[ \sum_{i=1}^{n} t_i^2 \right]^{1/2} \tag{22}$$

Total rise time of a digital optical link should not exceed 70% for a non return to zero code (NRZ) bit period, and 35% of a return to zero (RZ) code bit period. Assuming both transmitter and receiver as first order low pass filters, the transmitter and receiver rise times are given by:

$$t_{tx} = t_{rx} = \frac{350}{B_{tx}} = \frac{350}{B_{rx}}, \; n\sec \tag{23}$$

where $B_{tx}$ and $B_{tx}$ are the transmitter and receiver bandwidths in MHz. The bandwidth $B_M$ (L) due to modal dispersion of a digital optical link length L is empirically given by:

$$B_M(L) = \frac{B_0}{L^q} \quad , \tag{24}$$

where $B_0$ is the bandwidth per Km (MHz-Km product) and $0.5 < q < 1$ is the modal equilibrium factor. Then the modal dispersion rise time is given by:

$$t_{\mathrm{mod}} = \frac{0.44}{B_M} = \frac{440\, L^q}{B_0}, \; n\sec \tag{25}$$

$$t_{GVD} = |D|\, L\, \sigma_\lambda \;, n\sec \tag{26}$$

where D is the chromatic dispersion parameter (nsec/nm.km), $\sigma_\lambda$ is the half power spectral width of the source (nm), and L is the optical link distance in Km. Therefore the total rise time system is given by [15]:

$$t_{sys} = \left[ t_{tx}^2 + t_{rx}^2 + D^2 \sigma_\lambda^2 L^2 + (440)^2 L^{2q} / B_0^2 \right]^{1/2}, n\sec \tag{27}$$

# IV. RESULTS AND DISCUSSIONS

## *IV. 1. Wireless optical link*

The main objective of the wireless optical link design is to get as much light as possible from one end to the other, in order to light as possible from one end to the other, in order to receive a stronger signal that would result in higher link receive a stronger signal that would result in higher link margin and greater link availability. As shown in Table 1, the proposed wireless optical link parameters to achieve maximum both tranmission link distance and transmission data rate.

TABLE 1. PROPOSED WIRELESS OPTICAL LINK DESIGN PARAMETERS.

| | |
|---|---|
| Power transmitted ($P_T$) | 100 mWatt |
| Operating wavelength range (λ) | 0.85 μm to 1. 55 μm |
| Transmitter beam diveregnce (θ) | 115 degree |
| Recriver diameter ($D_R$) | 0.1-0.5 m |
| Link distance range | 0.1 to 10 Km |
| Receiver sensitivity ($S_R$) or power received | 2 μWatt |
| Transmitter and receiver losses (η) | 50 % |

Based on the assumed set of the controlling parameters for wireless optical link design to achieve the best transmission bit rates and transmission distances and the set of the figures from (5-10), the following facts are assured:

1) Fig. 5 has indicated that as the transmission distance (visibility) increases, the signal attenuation decreases at the same optical signal wavelength. While as the optical signal wavelength increases, signal attenuation decreases at the same transmission distance.

2) As shown in Fig. 6, as the beam diameter at receiver increases, the ray losses also increases at the same lens diameter. While as the lens diameter increases, the ray losses decrease at the same beam diameter at receiver.

3) Fig. 7 has demonstrated that as wireless optical link distance increases, the optical signal to noise ratio (OSNR) decreases at the same optical signal wavelength. Moreover, as the optical signal wavelength increases, the OSNR also increases at the same wireless optical link distance.

4) As shown in Fig. 8, as the transmitted radio frequency increases, the signal transmission also increases for both amplification and non amplification techniques. But with amplification technique offered high signal transmission.

5) Figs. (9, 10) have indicated that as the transmitted radio frequency increases, the transmission data rate also increasesin both cases of amplification and non amplification techniques at the same wireless link distance. While, as the wireless link distance increases, the transmission data rate decreases at the same transmitted radio frequency. Moreover with amplification techniques offered both high transmission link diatance and transmission data rate.

### *IV. 2. Wire optical cable link*

The main goal is how to develop a simple point to point digital wire optical cable link design, taking into account link power budget calculations and link rise time calculations. A link should satisfy both these budgets such as transmission distances, and data rate for a given BER. The data transmission bit rate, and transmission distances are the major factors of our interest for designing digital wire optical cable link. Table 2 shows the proposed wire digital optical cable link parameters to calculate both transmission distances and date rates.

TABLE 2. PROPOSED DIGITAL WIRE OPTICAL CABLE LINK DESIGN PARAMETERS.

| | |
|---|---|
| Power transmitted ($P_T$) | 100 mWatt |
| Power received ($P_{receiver}$) | 2 μWatt |
| Fiber Loss | 3.5 dB/Km |
| Couplers [LED-PIN] | 1.5 dB |
| Bandwidth per Km ($B_0$) | 900 MHz-Km |
| Modal equilibrium factor (q) | 0.7 |
| LED [$\sigma_\lambda$] | 50 nm |
| LD [$\sigma_\lambda$] | 1 nm |
| Couplers [LD-APD] | 8 dB |
| Material dispersion ($D_{mat}$) | 0.07 nsec/nm.Km |

Also in the same way, based on the assumed set of the controlling parameters for wire optical cable link design to achieve the best transmission bit rates and transmission distances and the set of the figures from (11-14), the following facts are assured:

6) As shown in Fig. 11, as the transmission data rate increases, the transmission distance decreases at the same optical signal wavelength. Moreover as the optical signal wavelength increases, the transmission distance also increases at the same transmission data rate.

7) Figs. (12, 13) have demonstrated that as the transmission data rate increases, the transmission distance decreases at the same attenuation limit for both LD/APD and LED/PIN for both single and multi-mode fibers. While as the attenuation limit with both LD/APD and LED/PIN decreases, the transmission distanceincreases at the same transmission data rate.

8) Fig. 14 has assured that as the transmission data rate increases, the transmission distance decreases at the same dispersion limit for both return to zero (RZ) and non return to zero (NRZ) codes. Moreover as the dispersion limit for both RZ, and NRZ decrease, the transmission distance increases at the same transmission data rate for single mode fiber link.

## V. CONCLUSIONS

In a summary, we have investigated and analyzed the transmission performance characteristics for both digital wire and wireless optical links in local and wide areas optical networks. We have demonstrated that the larger of the optical signal wavelength, the higher transmission distance for both wireless and wire digital optical links. Moreover, we have demonstrated that with amplification techniques, which added additional costs for wireless system, the wireless optical link offered both high transmission distances and transmission data rate. In the normal case (without amplification), the digital wire optical cable link offered both high transmission distances and data rates over wireless optical link with amplification. Also, we have assured that the use of LD/APD couplers offered maximum transmission distances and data rates over the use of LED/PIN couplers. Therefore it is evident that the digital wire optical cable links offered the best performance in cost, transmission distances, and transmission data rates over wireless optical links.

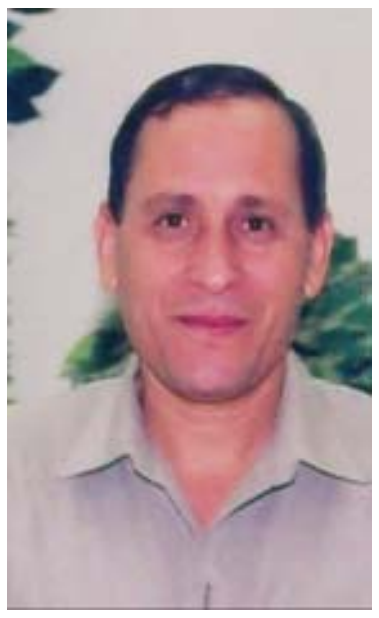

**Abd-Elnaser A. Mohammed**

Received Ph.D scientific degree from the faculty of Electronic Engineering, Menoufia University in 1994. Now, his job career is Assoc. Prof. Dr. in Electronics and Electrical Communication Engineering department. Currently, his field and research interest in the all passive optical and communication Networks, analog-digital communication systems, optical systems, and advanced optical communication networks.

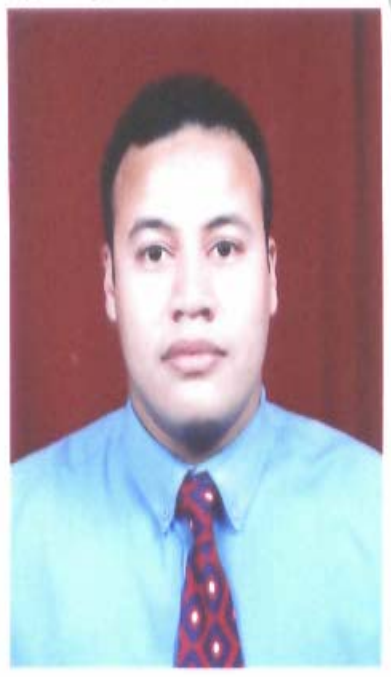

**Ahmed Nabih Zaki Rashed**

was born in Menouf, Menoufia State, Egypt, in 1976. Received the B.Sc. and M.Sc. practical scientific degrees in the Electronics and Electrical Communication Engineering Department from Faculty of Electronic Engineering, Menoufia University in 1999 and 2005, respectively. Currently, his field interest and working toward the Ph.D degree in Active and Passive Optical Networks (PONs). His theoretical and practical scientific research mainly focuses on the transmission data rates and distance of optical access networks.